\documentclass[pra,bibnotes,twocolumn]{revtex4}
\usepackage{amsmath}

\usepackage{graphicx}
\usepackage{hyperref}
\usepackage{graphicx}
\usepackage{amsmath,amssymb}
\usepackage{float}
\usepackage{bm}

\begin{document}
\draft

\def\ds{\displaystyle}
\title{ Quantum inverted harmonic potential}
\author{C. Yuce}
\address{Department of Physics, Faculty of Science, Eskisehir Technical University, Turkey }
\email{cyuce@eskisehir.edu.tr}
\date{\today}

\begin{abstract}
We consider a non-interacting gas under the inverted harmonic potential and present infinitely degenerate non-stationary orthogonal states. We discuss that it has an infinite entropy at the absolute zero temperature. We show that uncertainty in position of a particle under the inverted harmonic potential can be zero as there exists a solution which asymptotes to a Dirac delta function. We obtain a new free particle wave packet using the eigenstates for the inverted harmonic potential. It has unique self-focusing feature and can be used as focusing beam  without a lens in optical systems where paraxial approximation is used.
\end{abstract}

\maketitle

\section{Introduction}

The harmonic oscillator problem in quantum theory is an exactly solvable problem and can be used to model various physical systems. The inverted harmonic potential, a less known parabolic potential, has attracted some attention over the years \cite{iv7,iv8,iv9,iv10,iv11,iv12,iv13,iv14,iv15,iv15vs,cy01,cy02} because it has a wide range of application in many branches of physics. For example, an LC circuit with negative inductance and capacitance in quantum mesoscopic system is described by a model with an effective inverted harmonic potential \cite{iv0}. The inverted harmonic potential is also used in the study of fast frictionless cooling of ultracold atomic mixtures \cite{iv1,iv2,iv3} and and light propagation in inhomogeneous media \cite{iv5}. In condensed matter, transition layer states in type-I and type-II Weyl semimetals are governed by an effective inverted harmonic potential \cite{iv4}. The Hamiltonian for the inverted harmonic potential arises in conduction bands for the strained semiconducting transition metal dichalcogenides \cite{iv01} and in the theory of Hawking-Unruh Effect in Quantum Hall Systems \cite{iv6,iv6a,iv6b,iv6c}. Furthermore, an effective Hamiltonian with the inverted harmonic potential is used to study tachyonic backgrounds in 2D string theory \cite{ivhd1}, the quantum states of a D0-brane \cite{ivhd2}, quantum mechanics of the scalar field in the inflationary universe scenario \cite{ivhd3}, thermal activation problem \cite{ivhd4} and statistical fluctuations of fission dynamics  \cite{ivhd8} and to model instability in open quantum systems \cite{ivhd5,ivhd6,ivhd7}.\\
Although harmonic oscillator problem is well known and studied in every standard quantum mechanics text book, the quantum mechanical system with inverted harmonic potential has not been fully understood yet. In this paper, we study inverted harmonic potential problem and obtain non-stationary orthogonal states analytically. We show that they are infinitely degenerate with zero energy. This implies that entropy can be infinite in a system of non-interacting particles under inverted harmonic potential. We construct non-stationary ladder operators for the inverted harmonic potential. We find non-square integrable eigenstates for inverted harmonic potential and use them to calculate energy eigenvalues for a trapped particle under inverted harmonic potential. We further apply them to find bounded eigenstates in a semi-infinite system. We show that finite number of bounded eigenstates appear in a semi-infinite system with an inverted harmonic potential well. We discuss that these finite number of eigenstates are square integrable but can have infinite uncertainties. We finally obtain a new free particle wave packet using the inverted harmonic potential eigenstates with certain restrictions. The new wave packets have unique self-focusing feature and can be used as focusing beam where paraxial wave equation is used instead of the Schrodinger equation.

\section{Non-stationary orthogonal solutions}

Consider the Hamiltonian for 1D inverted harmonic potential
\begin{equation}\label{rekz902} 
 \mathcal{H} =\frac{p^2}{2m}-\frac{m~\omega^2}{2} x^2
\end{equation} 
where $m$ and $\ds{\omega}$ are constants. The inverted harmonic potential energy is not bounded from below, i.e., it goes to $\ds{ -  \infty  }$ as $\ds{ x \rightarrow  \mp \infty}$. Therefore, we intuitively expect that square integrable bounded eigenfunctions on the line $\ds{ (-\infty,\infty) }$ are not available. In other words, an arbitrary square integrable wave packet spreads out unboundedly since there is no minimum of the potential energy function. In this paper, we look for orthogonal set of square integrable {\it{non-stationary}}  wave packets for this problem. \\
Let $\ds{\Psi_n(x,t) }$ be a set of non-stationary solutions of the time-dependent Schrodinger equation $\ds{ \mathcal{H}\Psi_n(x,t )=i\hbar ~\partial_t  \Psi_n(x,t )   }$, where $n=0,1,2,...$ are non-negative integers. We demand that they are square integrable and orthogonal to each other, i. e., $\ds{  <\Psi_n(x,t)|    \Psi_{n^{\prime}} (x,t)>=\delta_{n,n^{\prime}} }   $, and $\ds{ \Psi_n(x \rightarrow  \mp \infty)=0}$. Note that $\ds{\Psi_n(x,t) }$ are not eigenstates but non-stationary states. We expect that $\ds{\Psi_n(x,t) }$ can be expressed in terms of the standard harmonic oscillator eigenstates since any square integrable and orthogonal wave packets on the line $\ds{ (-\infty,\infty) }$ can in principle be written in terms of the harmonic oscillator eigenstates, which are given by $\ds{ \phi_n(x)=N_n \exp{(-\frac{m{\Omega}x^2}{2\hbar})} ~H_n( \sqrt{\frac{m\Omega}{\hbar}}  x) }$, where $\ds{\Omega}$ is a constant, $\ds{N_n}$ are normalization constants and $\ds{H_n}$ are Hermite polynomials. For the inverted harmonic potential, this equivalence take a simple form and $\ds{ \Psi_n(x,t)}$ is given by
\begin{equation}\label{jhfwkl018d} 
 \Psi_n(x,t)=  \frac{ \exp \left( i\frac{m }{2\hbar  }~ \frac{  \dot{L} }{L} ~x^2  -i   (n+\frac{1}{2}) ~ {\Omega} ~ \tau   \right)  }{ \sqrt{L} }~  \phi_n(   \frac{x}{L} ) 
\end{equation} 
where the scale function $L(t)$ describes how the width of the non-stationary wave packet changes in time through the equation $\ds {\ddot{L} =\omega^2L+\frac{ \Omega^2 }{L^3}}$ and dot indicates a derivative respect to time $\ds{t}$ and $\ds{\frac{d\tau}{dt}=L^{-2} }$. Note that  $\ds{\Omega}$ is a constant which can be chosen freely. For the sake of simplicity, we set $\ds{\Omega=\omega}$. One can then analytically solve the equation for $\ds{L(t)  }$ with the initial conditions $\ds{L(0) =1}$ and $\ds{\dot{L}(0) =0}$ for which the initial wave packet $\ds{\Psi_n(x,0)}$ exactly matches the eigenstates of the harmonic oscillator. The corresponding solution reads $\ds{L(t)=\sqrt{\cosh{(2 \omega  t)} ~}  }$, which increases exponentially in time and the time parameters satisfy the relation $\ds{ \tan(\omega\tau)=  \tanh (\omega {t})  } $. This is in agreement with our expectation that such square-integrable wave packets tend to expand unboundedly since the potential is not bounded below. Fortunately, no transition among the non-stationary states occurs as they expand. For example, if the system is initially in the ground non-stationary state $\ds{n=0}$, then the system will be at the same state at all other times. Consider next the initial conditions $\ds{L(0) =1}$ and $\ds{\dot{L}(0) \neq0}$, where we assume a complex phase is  imprinted to the initial wave function. If $\ds{\dot{L}(0) <0}$, the initial phase forces to contract the wave packet up to a critical length and then the repulsive potential takes over and the wave packet starts to spread unboundedly.\\
One can easily see that the normalization condition is automatically satisfied for the non-stationary states; $\ds{ \int_{-\infty}^{\infty}   |\Psi_n(x,t)|^2 dx=  \int_{-\infty}^{\infty}   |\phi_n(z)|^2 dz }$, where $z=x/L$. They are also orthogonal$\ds{ \int_{-\infty}^{\infty}   \Psi^{\star}_m \Psi_n ~dx= 0 }$ for $\ds{m{\neq}n}$. Therefore an arbitrary square integrable initial wave packet can be expanded in terms of the non-stationary states (\ref{jhfwkl018d}) and its time evolution can then be easily studied. To this end, we stress that a different initial condition $\ds{\dot{L}(0) }$ leads to another set of orthogonal non-stationary states. Here the non-stationary states with $\ds{L(t)=\sqrt{\cosh{(2 \omega   t)} ~}  }$ will be our reference solutions and we note that any other set of orthogonal non-stationary solutions with $\ds{\dot{L}(0) \neq 0}$ can also be expanded in terms of our reference solutions. \\
The raising and lowering operators for the harmonic oscillator are given by $\ds{ \hat{a}^{\dagger} =  \sqrt{ \frac{m\omega}{{2\hbar}}} (~x- i\frac{p}{  m\omega    } )}$ and $\ds{ \hat{a} = \sqrt{ \frac{m\omega}{{2\hbar}}}  (   ~x+i\frac{p}{  m\omega    } )}$, respectively. A question arises. Do such ladder operators exist also for our non-stationary states? The answer is positive. Let us introduce time-dependent raising and lowering operators
\begin{eqnarray}\label{jhpokjhvd} 
\hat{A}^{\dagger} &=&   \sqrt{ \frac{m\omega}{{2\hbar}}}   \left(  ~(1 +   i   \frac{ L \dot{L}   }{  \omega  }   )\frac{x}{L}   -i  L \frac{  p  }{   m\omega   } \right)\nonumber\\
\hat{A} ~&=&  \sqrt{ \frac{m\omega}{{2\hbar}}}   \left(   ~(1 -  i   \frac{ L \dot{L}   }{  \omega  }   )\frac{x}{L}   +  i  L \frac{  p  }{   m\omega    }  \right) 
\end{eqnarray} 
where they operate on the non-stationary states: $\ds{   \hat{A} ~ \psi_n=\sqrt{n} ~  \psi_{n-1}}$ and $\ds{   \hat{A}^{\dagger}   ~\psi_n=\sqrt{n+1} ~  \psi_{n+1}}$, where $\ds{  \Psi_n= e^{ -i   (n+\frac{1}{2}) ~ {\Omega}  \tau  }  ~   \psi_{n}}$. Note that $\ds{  [ \hat{A}  ,\hat{A}^{\dagger}   ]=1  }$. One can produce an infinite set of non-stationary states by acting on $\psi_0$ with the operator  $\ds{ \hat{A}^{\dagger}  }$. These new ladder operators can be useful to find various interesting phenomena in a simple way. For example, one can construct non-stationary coherent states for our system by using the equation $\ds{   \hat{A}  \psi_n=\alpha ~  \psi_{n}}$, where $\alpha$ is a constant. \\
Let us now obtain the spectrum by computing $\ds{E_n=<   \Psi_n| \mathcal{H}  \Psi_n >}$. After a lengthy, but straightforward calculation, we get the energy values for each non-stationary states
\begin{equation}\label{ejkz018dilla} 
E_n=(\frac{1}{2L_0^2}-\frac{L_0^2}{2}+\frac{  \dot{L}_0^2  }{2\omega^2}  )~ (n+\frac{1}{2}) ~ \hbar ~\omega
\end{equation} 
where $\ds{L_0}$ and $\ds{\dot{L}_0}$ are the initial values of $\ds{L(t)}$ and $\ds{\dot{L}(t)}$, respectively. Note that $\ds{E_n}$ are time-independent. We stress that energy of a linear combination of the non-stationary states can not be calculated simply as in stationary eigenstates, i. e., $\ds{<\Psi  | \mathcal{H} \Psi>    \neq \sum_n  |a_n|^2 E_n }$, where $\ds{\Psi= \sum_n a_n \Psi_n}$  and $a_n$ are constant coefficients ($\ds{\Psi_n}$ are not eigenstates of $\ds{ \mathcal{H} }$ even if they are orthogonal to each other). Let us assume the initial conditions $\ds{ L_0=1   }$ and $\ds{  \dot{L}_0=c~\omega   }$, where $\ds{c}$ is a real-valued constant. Then, Eq. (\ref{ejkz018dilla}) becomes
\begin{equation}\label{wkl018dims} 
E_n=\frac{  c^2  }{2}  ~ (n+\frac{1}{2}) ~ \hbar~ \omega
\end{equation} 
The energy gap between two adjacent non-stationary states is equal to $\ds{ \frac{  c^2  }{2} \hbar \omega  }$. The striking point here is that $\ds{\Psi_n}$ become degenerate when $\ds{c=0}$
\begin{equation}\label{wcemyuce77} 
E_n=0~,~~~~for~ all ~n
\end{equation} 
This means that our reference orthogonal set of non-stationary states with $\ds{L(t)=\sqrt{\cosh{(2 \omega  t)} ~}  }$ (where $\ds{ L_0=1   }$ and $\ds{  \dot{L}_0=0  }$) have all zero energy. The kinetic energy is always positive and the potential energy is always negative for our system. The kinetic energy of a non-stationary state increases with increasing quantum number $\ds{n}$. On the other hand, the corresponding potential energy decreases with increasing quantum number $\ds{n}$ since the non-stationary waves become more extended. At a special value $\ds{c=0}$, they cancel each other and the total energy becomes zero. However, it is totally surprising that the cancellation occurs for all $\ds{n}$ and the system peculiarly becomes infinitely degenerate.

\subsection{  Infinite degeneracy }

This infinite degeneracy has some interesting consequences, such as infinite entropy.  Below, we will qualitatively discuss this issue. \\
Consider a non-interacting gas under inverted harmonic potential. To study the system, we use the orthogonal non-stationary states that are statistically independent from each other. Suppose that the system is isolated, i. e., no particle and energy transfer to and from its surroundings are allowed. Suppose further that the system is at absolute zero temperature $\ds{T=0}$ and the internal energy is zero, $\ds{U=0}$. In this case, we start with $\ds{\Psi_n}$ with $\ds{c=0}$. Since the ground state is infinite-times degenerate, the entropy of the system is infinite (  it is uncertain which state the particle occupies as all states have zero energy. There are infinitely many ways to distribute all particles among the non-stationary states $\Psi_n$ with $\ds{c=0}$. Recall that entropy is a measure of the number of possible ways energy can be distributed in a system). The concept of infinite entropy is true for both fermions and bosons. But this is not consistent with the idea that entropy should be a well-defined constant at absolute zero. Note that the entropy is infinite no matter how many particles there are in the system. It is well known that entropy is an extensive property of a thermodynamic system, which means that its value changes with the total number of particles. But, this is not a paradox since the entropy is infinite in our system. Suppose next that the internal energy of the system is not zero $\ds{U\neq0}$. In this case, we can use another set of non-stationary orthogonal states to study our system. For example, one can fix $\ds{c=1}$ (Recall that a non-stationary state with $c\neq0$ can be expanded in terms of the non-stationary states with $c=0$). According to Eq. (\ref{wkl018dims}), the energy spectrum is no longer infinitely degenerate and hence we say that the corresponding entropy becomes finite. In other words, the entropy is infinite at $T=0$ and the heat transfer into the system decreases the entropy by an infinite amount.\\

\section{ An integral representation}

Having established the equivalence between the systems with harmonic and inverted harmonic potentials, we are now in position to explore another equivalence between the inverted harmonic oscillator and free particle. Let us now set $\ds{ \Omega=0  }$. In this case, the solutions of the equation $\ds{\ddot{L} =\omega^2 L }   $ are given by $\ds{L(t)  =L_+e^{    {\omega}t}  +L_-e^{    - {\omega}t}    }$, where $\ds{ L_{\mp}}$ are arbitrary coefficients. Instead of the harmonic oscillator eigenstates in (\ref{jhfwkl018d}), we use the most general  free particle wave function in $\ds{k}$-space. If we substutite $\ds{\phi(\frac{x}{L})  =  \frac{1}{\sqrt{2\pi\hbar}}  \int_{-\infty}^{\infty}  e^{i (\frac{kx}{ L} -\frac{E}{\hbar} \tau)} ~\psi(k)~dk  }$ in (\ref{jhfwkl018d}), then we get the most general non-stationary solutions for the inverted harmonic potential
\begin{equation}\label{jhf18dyr4u} 
 \Psi   =  \frac{1}{\sqrt{2\pi\hbar}}  \frac{    e^{ i\frac{m\omega^2    \dot{L} }{2\hbar~L   }~ x^2     }}{ \sqrt{  L } }  \int_{-\infty}^{\infty} e^{i (\frac{kx}{ L} -\frac{E}{\hbar} \tau)} ~\psi(k)~dk
\end{equation} 
where $\ds{\psi(k) }$ is an arbitrary function, $\ds{E=\frac{\hbar^2}{2m} k^2}$ is a continuous parameter, which does not match the energy values for this wave packet, $\ds{  \frac{d \tau}{dt}=\frac{  1 }{L^2} }$ and the integral is over k-space. The energy of such a wave packet can be calculated once $\ds{\psi(k)}$ is specified. This equation allows us to find time evolution of a wave packet (square integrable or not) starting from its corresponding function $\ds{\psi(k) }$. Note that $\ds{\psi(k) }$ matches the $k$-space wave packet of $\ds{\Psi }$ at $t=0$ when $\ds{L_-=L_+=0.5}$.  The case with $\ds{L_-=1}$ and $\ds{L_+=0}$ is also interesting since one can construct shrinking wave packets (whose widths decrease in time).  As an example, consider the initial truncated Airy wave packet $\ds{\phi(x,t=0)=  e^{    \alpha{  (\frac{x}{L}-\frac{\tau^2}{2} )}   }  Ai(x) }$, where $Ai(x)$ is the Airy function \cite{Air1,Air2}. Then the absolute value of the wave packet evolves according to
$
 \left | \Psi   \right| =\frac{  e^{    \alpha{  (\frac{x}{L}-\frac{\tau^2}{2} )}   }  }{\sqrt{L}} \left | Ai(\frac{x}{L}-\frac{\tau^2}{2} +i{ \alpha} \tau ) \right|
$, where $\ds{L=e^{-{\omega}t}}$, $\ds{\tau=\frac{e^{ 2 \omega {t} }-1}{\omega } }$, $m=\hbar=1$ and $\ds{\alpha>0   }$ is the decay factor to make the wave packet square integrable. The initial quadratic phase in $\ds{ \Psi}$ gives extra kinetic energy to the initial wave packet. Furthermore, the minus sign of the initial quadratic phase implies that the wave packet is initially contracting. The potential is repulsive and hence decelerates this contraction but the wave packet's initial kinetic energy is big enough not to be stopped. In other words, its width decreases with a slowing rate and shrinks into a singular point while its peak intensity grows (it becomes like a Dirac delta function ) at $\ds{t\rightarrow \infty}$. This implies that the position of the particle can become certain. This does not contradict with the Heisenberg uncertainty principle as it happens at infinite time. This behavior is analogous to the classical behavior. The solution for the 1D classical particle under the inverted harmonic potential is given by $\ds{  x(t)=x_0 e^{\mp {\omega}t}    }$, where $\ds{x_0}$ is the initial position and $\ds{v_0=\mp{\omega}x_0}$ is the corresponding initial velocity. If $\ds{v_0>0}$, then the particle moves unboundedly since there is no minimum of the potential. If $\ds{v_0<0}$, then the particle decelerates towards the center. Despite the potential is repulsive, the initial kinetic energy is big enough not to be stopped until it reaches the center at $\ds{t\rightarrow \infty}$. 

\section{Stationary solutions}

In this section, we obtain the eigenstates for the inverted harmonic potential. They are not square integrable and don't decay at infinities since the potential is not bounded from below. \\
The solutions of the time-independent Schrodinger equation with parabolic potential are known in the literature \cite{cy02}. The two linearly independent eigenstates for the Hamiltonian (\ref{rekz902}) can be given in terms of the parabolic cylinder functions $\ds{D_\nu(x) }$
\begin{eqnarray}\label{r72e84ejjgfgd} 
\psi_1(x)&=&D_{   \frac{iE}{\hbar \omega}-\frac{1}{2} }   ( i  \sqrt{ \frac{ 2i m\omega}{\hbar} } x )   \nonumber\\
\psi_2(x)&=& D_{ -  \frac{iE}{\hbar \omega}-\frac{1}{2} }   (   \sqrt{ \frac{ 2i m\omega}{\hbar} } x )    
\end{eqnarray}
where the parameter $\ds{E}$ describe the continuous energy eigenvalues. $\ds{\psi_1}$ oscillates as $\ds{x\rightarrow \infty}$ and vanishes as $\ds{x\rightarrow -\infty}$ and vice versa for $\ds{\psi_2}$. Therefore they  are not square integrable. One can also express them
in terms of the hypergeometric functions. The even and odd solutions are given by  \cite{ablos}
\begin{eqnarray}\label{03eweye?8lsd} 
\psi_{odd}&=& x~ e^{-i \frac{ m\omega}{2\hbar } x^2  }  ~F (   \frac{iE}{2\hbar \omega}+\frac{3}{4}    ,\frac{3}{2}, i   \frac{ m\omega}{\hbar}  x^2 )\nonumber\\
\psi_{even}&=&~~e^{-i \frac{ m\omega}{2\hbar } x^2  }  ~~ F (   \frac{iE}{2\hbar \omega}+\frac{1}{4}    ,\frac{1}{2}, i   \frac{ m\omega}{\hbar}  x^2 ) 
\end{eqnarray}
where $F$ is the confluent hypergeometric function. The most general solution is a superposition of these eigenfunctions. Below, we use them to find exact solutions for some problems such as trapped inverted oscillator and parabolic potential well. 

\subsection{A trapped particle under the inverted harmonic potential}

Consider a particle under the inverted harmonic potential. The particle is also trapped in a box of width $\ds{d=d^{\prime} \sqrt{ \frac{\hbar}{m\omega}}}$ with rigid walls. Therefore we solve the Schrodinger equation for the inverted harmonic potential subject to the boundary conditions  $\ds{\Psi(x=0)=\Psi( x=d)=0}$.\\
Let us firstly discuss this problem qualitatively. The kinetic and potential energies for the trapped particle scales with $\ds{E_k\sim \frac{1}{d^2}}$ and $\ds{E_p\sim-d^2}$, respectively. The kinetic (potential) energy is dominant for a sufficiently small (large) box. Therefore, we can say that the ground state energy is positive for a small box and becomes exactly zero at a critical length $\ds{d_c  }$ (we numerically check this argument and find $\ds{d^{\prime}_c=2.36  }$). Beyond this critical length, the ground state energy becomes negative. We claim that we can predict energy eigenvalues for a small box using the energy eigenvalues for a free particle in an infinite box as long as $\ds{  E_{n=0}^{free}  >> |V_{min}   |}$, where $\ds{  E_{n=0}^{free} }$ is the ground state energy for the free particle and $\ds{V_{min}=- \frac{  ~ {d^{\prime}  }^2}{2}  {\hbar \omega} }$ is the minimum potential in the box. Therefore $\ds{\frac{E_n}{\hbar \omega}\approx        \frac{n^2 \pi^2}{2  {d^{\prime}}^2}    =\frac{1}{  {d^{\prime}}^2  }\{4.935, 19.739, 44.413, 78.957,...\}  }$ when $\ds{d^{\prime} <1  } $. \\
Let us now solve our problem exactly by applying the boundary conditions: $\ds{\Psi(x=0)=\Psi( x=d)=0}$ where $\ds{\Psi=a_1\psi_{odd}+a_2\psi_{even}} $ and $a_1$ and $a_2$ are constants to be determined from the boundary conditions and $\ds{\psi_{odd,even}}$ are given in (\ref{03eweye?8lsd}). It is easy to see that $a_2=0$ since $\psi_{even}$ is not zero at $\ds{x=0}$. Let us apply the boundary condition at the right wall. Then this problem reduces to searching for zeros of the confluent hypergeometric function and can be easily solved using numerical methods. We numerically calculate the first 4-lying energy eigenvalues when $\ds{d^{\prime}=1}$. They are given by $\ds{   \frac{E}{\hbar \omega}=\{4.793, 19.579, 44.250, 78.791     ,...\}}$. These are in good agreement with our above derivation using the free particle energy eigenvalues. If $\ds{d^{\prime}}$ increases, the deviation from the free particle energy eigenvalues becomes significant. For example, $\ds{   \frac{E_n}{\hbar \omega}=\{0.632, 4.305, 10.454,19.082      ,...\}}$ at $\ds{d^{\prime}=2}$ and $\ds{   \frac{E_n}{\hbar \omega}=\{-1.151, 0.883,  3.535, 7.332     ,...\}}$ at $\ds{d^{\prime}=3}$.   

\subsection{Finite number of bounded eigenstates }

Let us now study a semi-infinite system for the inverted harmonic potential $\ds{V=-m\omega^2x^2/2}$. Suppose that an infinitely hard wall is placed at $\ds{x=-d}$ so that a potential well is formed for $\ds{ x\in [-d,0]  }$. The local minimum of the potential energy occurs at $x=-d$ while the potential is repulsive unboundedly when $x>0$. Therefore, we intuitively expect that both bounded and unbounded solutions can be simultaneously available for this problem. \\
To solve the problem analytically, we use the eigenstates given in Eq. (\ref{r72e84ejjgfgd}) and set the boundary conditions
\begin{equation}\label{r3892m1lzs} 
\Psi(x=-d)=\Psi( x=\infty)=0
\end{equation} 
The right boundary condition is easy to apply as $\ds{\psi_2}$ vanishes at $\ds{x\rightarrow \infty}$ while $\ds{\psi_1}$ does not. Then the left boundary condition allows us to find discrete energy eigenvalues: $ \psi_2(x=-d)= D_{ -  \frac{iE}{\hbar \omega}-\frac{1}{2} }   ( -  \sqrt{ 2i } ~d^{\prime} )    =0$ ($\ds{d=d^{\prime} \sqrt{ \frac{\hbar}{m\omega}}}$), which is satisfied only when energy eigenvalues are complex valued, $\ds{E=E_R +i E_I}$. We interpret that non-zero values of $\ds{E_I}$ lead to the leakage of the wave packet from the well into the positive $x$-region where the potential is not bounded. Fortunately, solutions with  $\ds{\frac{E_I}{\hbar \omega}<10^{-6}}$ can be practically considered as stationary bounded eigenstates. It is interesting to see that no such solutions exist until $\ds{d^{\prime  }}$ is less than a critical value. We numerically find that $\ds{d^{\prime  }=d^{\prime}_c\approx 3.51}$. In other words, the wave packet localized in the potential well is rapidly expelled into the positive $x$-region of the semi-infinite space as long as $\ds{d^{\prime  }<  d^{\prime}_c }$ . On the other hand, square integrable bounded eigenstates are available when $\ds{d^{\prime  }> d^{\prime}_c }$. Surprisingly, we find that there are finite number of such eigenstates and this number change discretely with $\ds{d^{\prime  } }$. More specifically, there exists only one bounded eigenstate with energy eigenvalue $\ds{E=-2.32 \hbar \omega}$ at $\ds{d^{\prime  }=  d^{\prime}_c }$. We think that this unique stationary eigenstate may find some practical applications. Increasing $\ds{d^{\prime  }}$ from $\ds{d^{\prime  }_c}$ does not change the number of available eigenstates but its energy eigenvalue until a second critical value, at which the number of available eigenstates becomes $2$. One can go in this way and concludes that the number of available eigenstates increases discretely by increasing $\ds{d^{\prime}}$. Unfortunately, these eigenstates do not decay exponentially fast, i. e.  $\ds{|\psi_2|^2 \sim \frac{1}{x}} $ as $\ds{x\rightarrow \infty}$. Therefore, $\ds{<x>}$ and $\ds{<x^2>}$ are infinite even if the eigenstates are square integrable. This is interesting as it is impossible to know where the particle is. In a real experiment, the inverted harmonic oscillator does not extend up to infinity, so uncertainties $\ds{{\Delta}x}$ take finite values.

\section{Self-focusing free particle solutions}

In Section III, we have obtained a general solution for the inverted harmonic potential using free particle solution. Here, we study the problem the other way around. In other words, we obtain a new free particle solution using the eigenstates of the inverted harmonic potential (\ref{03eweye?8lsd}). This new solution is not square integrable like plane waves and has an interesting self-focusing character.\\ 
Let us set $\ds{\omega=0}$ and $\ds{\Omega^2=-\omega^2}$ and use the ansatz  $    \Psi =  \frac{ \exp \left( i\frac{m }{2\hbar  }~ \frac{  \dot{L} }{L} ~x^2  -i  \frac{E}{\hbar} ~ \tau   \right)  }{ \sqrt{L} }~  \phi(   \frac{x}{L} ) 
$ (\ref{jhfwkl018d}) and substitute either $\ds{\phi( \frac{x}{L})  =\psi_{odd}( \frac{x}{L})   }$ or $\ds{\phi( \frac{x}{L})  =\psi_{even}( \frac{x}{L})   }$ (\ref{03eweye?8lsd}). Therefore, we obtain two non-stationary and non-square integrable odd and even solutions for free particle
\begin{eqnarray}\label{jhfls57849eowlsd} 
\Psi_{E}^-&=&  \frac{  x~ e^{ -  i \frac{m { \omega }B_F ~    x^2 }{2~\hbar    }  - i\frac{E }{\hbar} \tau  } }{  \sqrt{L_F^3}  } 
 F (   \frac{iE}{2\hbar \omega}+\frac{3}{4}    ,\frac{3}{2}, i   \frac{ m\omega}{\hbar}  \frac{x^2}{L_F^2} ) ~~  \nonumber\\
 \Psi_{E}^+  &=& \frac{  e^{ -  i \frac{m  {\omega}B_F  ~  x^2 }{2~\hbar    }  - i\frac{E }{\hbar} \tau  } }{  \sqrt{L_F}  }  F (   \frac{iE}{2\hbar \omega}+\frac{1}{4}    ,\frac{1}{2}, i   \frac{ m\omega}{\hbar}  \frac{x^2}{L_F^2} ) ~~~~
\end{eqnarray} 
where  $\ds{\ddot{L}_F= - \frac{\omega^2}{L_F^3}   }$, $\ds{ B_F=  \frac{\dot{L}_F}{L_F}- \frac{\omega}{L_F^2}       }$ and $\ds{\frac{d\tau}{dt}=L^{-2} }$. We solve the equation for $L_F(t)$ with the initial condition $L_F(0)=1$. Two specific solutions read $\ds{L_{F_1}=\sqrt{1-2{\omega} t}}$ and $\ds{L_{F_2}=\sqrt{1-{\omega}^2 t^2}}$ corresponding to $\ds{\dot{L}_F(0)=-\omega}$ and $\ds{\dot{L}_F(0)=0}$, respectively. \\
These new free particle solutions are interesting in the sense that their widths decrease continuously until the wave packets are all focused at $\ds{x=0}$ and then expand unboundedly. In other words, the system has a singularity at a critical time $\ds{   t_c =   \frac{1} {2\omega }     }$ ($\ds{   t_c =  \sqrt{   \frac{1} {\omega }  }   }$) at which $\ds{L_{F_1}}$ ($\ds{L_{F_2}}$) becomes zero. At the singular point, the width of the wave packet becomes zero. Note that the intensities of these wave packets grow to infinity when the widths become zero (like the Dirac-delta function). In this case, one may say that the position of the free particle can be exactly known at $\ds{t=t_{c}}$. It seems that it contradicts the Heisenberg uncertainty principle. But this is not a paradox as these two waves carry infinite energy since they are not square integrable ($\ds{ \Delta x   }$ and $\ds{ \Delta p   }$ are no longer physical as they diverge). In fact, these waves must be truncated to be experimentally realized. However, the singularity at $\ds{t_{c}}$ is lost whenever we truncate them. Fortunately, we can still exploit this focusing feature for such a truncated wave. In other words, the intensity of the main lobe of the truncated wave increases enormously up to a critical point and then decrease. \\
The above solutions may be used in the field of electron microscopy and especially in optics community where the paraxial wave equation (a Schrodinger-like equation) is used. In optics, self-focusing waves, which can be realized in nonlinear optical systems have some practical applications. Finding self-focusing light beam in free space is an important problem in optics community. Generally speaking, diffraction occurs when light beam is propagated in free space and the peak intensity value decreases during propagation. Our solution can be a candidate for self-focusing light beams in free space (beams focusing without a lens and diffracting rapidly beyond the focal point) and can be practically used in a variety of applications. 

\section{Conclusion}

The harmonic oscillator is of special importance in quantum theory as it models many quantum mechanical systems. Here we consider the Hamiltonian for the inverted harmonic potential and present some exact analytical solutions. We think our exact solutions can be used to model some systems in quantum theory. \\
It is well known that a system with a single available microstate has an entropy of zero. A perfect crystal at a temperature of absolute zero is the only such system known in physics. Oppositely, we predict here the existence of infinite entropy for a system of non-interacting particles under the inverted harmonic potential. To the best of our knowledge, it is the only system with an infinite entropy at $T=0$. This is the result of infinite degeneracy of non-stationary orthogonal states. Here, we also construct ladder operators for the non-stationary orthogonal states. \\
We present exact shrinking solutions for the inverted harmonic potential and explain them using classical arguments. We construct wave packets which asymptotes to a Dirac-delta function. \\
We obtain energy eigenvalues for a trapped particle under the inverted harmonic potential. We consider a semi-infinite system and show that a finite number of bounded eigenstates, which increases discretely with the size of the potential well, appear in the semi-infinite system. We discuss that they are square integrable but have infinite uncertainties. \\
We obtain a new free particle wave packet using the eigenstates for the inverted harmonic potential. The new wave packets have an infinite amount of energy and unique self-focusing feature. We think that this focusing solution without a lens can find applications in optics where paraxial wave approximation is used.


\begin{thebibliography}{0}
\bibitem{iv7} V. V. Chistyakov, Pramana -J. Phys. {\bf 91}, 57 (2018).
\bibitem{iv8} R. D. Mota, D. Ojeda-Guillen, M. Salazar-Ramirez and V. D. Granados, Mod. Phys. Lett. A  {\bf 34}, 1950028 (2019).
\bibitem{iv9} Amlan K. Roy, Mod. Phys. Lett. A  {\bf 30}, 1550176 (2015).
\bibitem{iv10} Soo-Chang Pei and Chun-Lin Liu, Journal of the Optical Society of America A {\bf 30}, 2096 (2013).
\bibitem{iv11} Ole Steuernagel, Eur. Phys. J. Plus . {\bf 129}, 114 (2014) .
\bibitem{iv12} I.A. Pedrosa, Alberes Lopes de Lima, and Alexandre M. de M. Carvalho, Can. J. Phys. {\bf 93}, 841 (2015).
\bibitem{iv13} Mustapha Maamache and Jeong Ryeol Choi, Chinese Phys. C {\bf 41}, 113106 (2017).
\bibitem{iv14} Carlos A Munoz, Juvenal Rueda-Paz and Kurt Bernardo Wolf, J. Phys. A: Math. Theor. {\bf 42}, 485210 (2009).
\bibitem{iv15} David Bermudez, David J. Fernandez C, Ann. Phys. {\bf 333}, 290  (2013).
\bibitem{iv15vs}V Subramanyan, SS Hegde, S Vishveshwara, arXiv:2012.09875 (2020).
\bibitem{cy01} C. Yuce, A. Kilic, A. Coruh, Physica Scripta {\bf 74}, 114 (2006).
\bibitem{cy02} C Yuce, Phys. Lett. A  {\bf 380}, 3791 (2016).
\bibitem{iv0} I. A. Pedrosa and E. Nogueira Jr., I. Guedes, Mod. Phys. Lett. A  {\bf 30}, 1650122 (2016).
\bibitem{iv1} Stephen Choi, Roberto Onofrio, and Bala Sundaram, Phys. Rev. A {\bf 84}, 051601(R) (2011).
\bibitem{iv2} Stephen Choi, Roberto Onofrio, and Bala Sundaram, Phys. Rev. A  {\bf 86}, 043436 (2012).
\bibitem{iv3} Stephen Choi, Roberto Onofrio, and Bala Sundaram, Phys. Rev. A {\bf 88}, 053401 (2013).
\bibitem{iv5} A. R. Urzua, et al. Sci. Rep. {\bf 9}, 16800 (2019).
\bibitem{iv4} Yu-Ge Chen, Xi Luo, Fei-Ye Li, Bin Chen, Yue Yu, Phys. Rev. B  {\bf 101}, 035130 (2020).
\bibitem{iv01} Habib Rostami, Rafael Roldan, Emmanuele Cappelluti, Reza Asgari, and Francisco Guinea, Phys. Rev. B  {\bf 92}, 195402 (2015).
\bibitem{iv6} Suraj S. Hegde, Varsha Subramanyan, Barry Bradlyn, and Smitha Vishveshwara, Phys. Rev. Lett.  {\bf 123}, 156802 (2019).
\bibitem{iv6a} Takeshi Morita, Eur. Phys. J. C {\bf 80}, 331 (2020).
\bibitem{iv6b} Takeshi Morita, Phys. Rev. Lett. {\bf 122}, 101603 (2019).
\bibitem{iv6c} Praloy Das, Subir Ghosh, arXiv:1905.00847 (2019).
\bibitem{ivhd1} S. Cremonini, JHEP {\bf 10}, 014 (2005).
\bibitem{ivhd2} J. Ambjorn and RA Janik, Phys. Lett. B {\bf 584}, 155 (2004).
\bibitem{ivhd3} A. H. Guth and S.Y. Pi, Phys. Rev. D {\bf 32}, 1899 (1985).
\bibitem{ivhd4}  Daniel Boyanovsky, Richard Holman, Da-Shin Lee, Joo P. Silva, Nucl.Phys. B {\bf 441}, 595 (1995).
\bibitem{ivhd8} H. Hofmann, D. Kiderlen, Phys. Rev. C  {\bf 56}, 1025 (1997).
\bibitem{ivhd5}  Paul A. Miller and Sarben Sarkar, Phys. Rev. E {\bf 58}, 4217 (1998).
\bibitem{ivhd6} F. H. Gaioli, E. T. GarciaAlvarez, M. A. Castagnino, Int. J. Theor. Phys.  {\bf 36}, 2371 (1997).
\bibitem{ivhd7} Wojciech Hubert Zurek and Juan Pablo Paz, Phys. Rev. Lett. {\bf 72} 2508 (1994).
\bibitem{Air1} G. A. Siviloglou, J. Broky, A. Dogariu, and D. N. Christodoulides, Phys. Rev. Lett.  {\bf 99}, 213901 (2007).
\bibitem{Air2} C. Yuce, Mod. Phys. Lett. B,  {\bf 29}, 1550171 (2015)
\bibitem{ablos} I. Stegun, M. Abramowitz, "Handbook of Mathematical Functions with Formulas, Graphs, and Mathematical Tables", over, New York, ninth Dover printing, tenth GPO printing edition, (1964).
\end{thebibliography}
 \end{document}